\documentclass[draft]{agujournal2019}
\usepackage{url} 
\usepackage{lineno}
\usepackage[inline]{trackchanges} 
\usepackage{soul}
\usepackage{booktabs}
\usepackage[ruled,vlined,linesnumbered]{algorithm2e}

\usepackage{amsmath, amssymb}
\usepackage{bm}

\SetKwInput{KwRequire}{Require}
\SetKwInput{KwEnsure}{Ensure}
\SetKwComment{Comment}{$\triangleright$ }{}

\draftfalse

\journalname{Journal of Advances in Modeling Earth Systems (JAMES)}

\begin{document}
	
	%
	%
	
	
	\title{A score-based particle flow filter for non-Gaussian data assimilation in high-dimensional chaotic systems}
	
	%
	%
	
	
	
	
	\authors{Zheqi Shen\affil{1}, Youmin Tang\affil{1,2} and Yuewei Fang\affil{1,3}}

	
	\affiliation{1}{College of Oceanography, Hohai University, Nanjing, China.}
	\affiliation{2}{Department of Geography, Earth and Environmental  Sciences, University of Northern British Columbia, Canada}
	\affiliation{3}{State Key Laboratory of Satellite Ocean Environment Dynamics, Second Institute of Oceanography, Ministry of Natural Resources, Hangzhou, China}

	
	
	
	\correspondingauthor{Youmin Tang}{ytang@hhu.edu.cn}

	
	
	\begin{keypoints}
		\item Score-PFF replaces Gaussian prior gradients with neural network-learned score functions to better characterize non-Gaussian attractor structures in chaotic systems
		\item The learned score captures multimodal and skewed prior distributions, yielding substantial RMSE reduction over Gaussian-based methods on the Lorenz-96 system
		\item  Neural network inference replaces SVD-based covariance inversion, eliminating the computational bottleneck of Gaussian PFF in high-dimensional systems

	\end{keypoints}
	
	%
	%
	
	%
	%
	
		
		\begin{abstract}
Current particle flow filters rely on Gaussian prior assumptions that fail to capture the non-Gaussian attractor structure of chaotic systems. This study proposes a Score-based Particle Flow Filter (Score-PFF) that replaces the parametric prior gradient with a neural network-learned score function via denoising score matching. This enables flexible characterization of multimodal, skewed, and complex prior distributions in chaotic dynamics.
Pure prior adjustment experiments demonstrate correct gradient directions toward the attractor (26.9\%–29.0\% error reduction over Gaussian priors). Under linear observations, Score-PFF significantly outperforms both Gaussian PFF and EAKF (Cohen's d  = 1.03 and 1.40), preserving non-Gaussian structure that EAKF progressively Gaussianizes. Under nonlinear observation operators, Score-PFF maintains robust performance with up to 60\% RMSE reduction in strongly non-Gaussian regimes.
On the 1000-dimensional Lorenz-96 system, Score-PFF achieves 49.5\% RMSE reduction over PFF while reducing per-assimilation cost by replacing SVD-based covariance inversion with neural network inference.
Score-PFF establishes a computationally tractable, non-Gaussian data assimilation framework suitable for high-dimensional geophysical systems.		
		\end{abstract}
		
	\section*{Plain Language Summary}
	Weather and ocean forecasting rely on combining computer models with real-world observations—a process called data assimilation. Traditional methods assume that forecast errors follow Gaussian (bell-shaped) distributions, which fails for complex chaotic systems like the atmosphere. Particle flow filters can handle non-Gaussian distributions but rely on expensive matrix calculations that become increasingly cumbersome in high dimensions.
	This study introduces Score-PFF, a new method that trains a neural network to learn the patterns of forecast errors from model simulations. Instead of performing slow matrix inversion operations, the trained network directly predicts how to adjust forecasts. Tests on a simplified atmospheric model show that this approach is more accurate than existing methods, especially when observations are nonlinear. The method also reduces computational bottlenecks by replacing matrix inversion with neural network inference, making it suitable for future operational weather forecasting systems.
	
			
		%
		%
		
		%
		
		
		%
		%
		%
		%

		\section{Introduction}\label{sec:intro}
		Data assimilation is a technique that combines numerical models with observational data to provide optimal estimates of initial conditions for numerical prediction, reduce uncertainties in model parameters, and generate reanalysis datasets to support scientific research and artificial intelligence model training. Traditional methods based on variational theory or sequential estimation---including three-dimensional variational (3D-Var), optimal interpolation (OI), and the Kalman filter (KF)---have been extensively studied and widely applied in operational systems. These methods heavily rely on linear numerical models and Gaussian assumptions. Variational methods such as 3D-Var and OI assume Gaussian prior, posterior, and likelihood functions; the KF assumes linear dynamics and Gaussian distributions, with the extended Kalman filter (EKF) introducing tangent linear approximations for weakly nonlinear systems that suffer from truncation errors. The ensemble Kalman filter (EnKF) addresses the computational limitations of EKF by employing Monte Carlo methods to avoid explicit storage of the background error covariance matrix while maintaining flow-dependent error evolution \cite{Evensen_2003}; however, it still relies on Gaussian assumptions for probability density functions. Under strongly nonlinear and non-Gaussian scenarios, the effectiveness of these methods is significantly limited.
		
		The particle filter (PF) provides a nonlinear assimilation method free of Gaussian assumptions \cite{VanLeeuwen_Hans2019}. Its fundamental idea is to approximate Bayesian filtering through importance sampling and resampling. However, PF suffers from severe particle degeneracy: in high-dimensional systems, most particle weights become vanishingly small, causing performance to deteriorate sharply. The probability of this phenomenon increases exponentially with dimensionality, making the "curse of dimensionality" unavoidable even with increased ensemble size \cite{Bengtsson_2008}.
		
		To alleviate this curse, researchers have proposed equal-weight particle filters, localized particle filters, and hybrid filters\cite{Snyder_2008, Poterjoy_2018, Robert_2017}. These improve high-dimensional performance through optimized proposal distributions, localization, or hybridizing EnKF and PF analysis steps \cite{Shen2015,Shen2017}. Nevertheless, such improvements still rely on resampling or Gaussian approximations and may face limitations in accuracy, parameter sensitivity, or computational efficiency under extreme non-Gaussian scenarios.
		
		The particle flow filter (PFF) offers a fundamentally different approach: it drives particles to evolve from prior to posterior through continuous-time differential equations, circumventing resampling and maintaining diversity in high dimensions. The theoretical foundation was established by \citeA{daum2011exact} with the Exact Daum-Huang Filter (EDH). Subsequent developments include Stein variational gradient descent \cite{liu2016stein} and the mapping particle filter \cite{pulido2019sequential}, which embed particle updates in reproducing kernel Hilbert spaces (RKHS). The current state-of-the-art in geophysical applications is the matrix-valued kernel PFF of \citeA{hu2021particle}, which achieves efficient assimilation in $10^5$--$10^6$ dimensional systems through locally adaptive bandwidth matrices and implementation within the Data Assimilation Research Testbed (DART) \cite{hu2024implementation}.
		
However, \citeA{hu2021particle} still relies on the Gaussian prior assumption, computing the prior gradient as $-\mathbf{B}_{\text{loc}}^{-1}(\mathbf{x}-\bar{\mathbf{x}})$. This linearization struggles to capture the non-Gaussian attractor structure of chaotic systems such as Lorenz-96. Moreover, the Gaussian parameterization necessitates explicit covariance inversion: $\mathcal{O}(n_x^3)$ for unlocalized matrices, or $\mathcal{O}(r n_x^2)$ with truncated SVD where $r$ is the effective rank (typically $r \sim 10$--$50$). This SVD-based computation suffers from numerical instabilities, sensitive tuning of localization parameters, and poor GPU utilization, creating practical bottlenecks for high-dimensional applications.
	
To address these limitations, this study proposes a score-based particle flow filter (Score-PFF): retaining the differential equation-driven framework of PFF, but replacing the Gaussian prior with score functions learned by deep neural networks \cite{vincent2011connection, song2020score}. This modification overcomes the fundamental limitation of characterizing non-Gaussian priors (multimodal, skewed distributions) inherent in chaotic dynamics. The neural network prior gradient also replaces the SVD-based covariance inversion required by Gaussian PFF, eliminating the dominant computational bottleneck that constrains scalability to high-dimensional operational systems.
	
Recently, diffusion models have emerged as an alternative for data assimilation, with approaches ranging from GraphCast-based post-processing \cite{huang2024diffda} to score-based state generation \cite{Rozet2023,yang2025gap,shen2025cdsm}. These methods demonstrate the representational power of score functions for geophysical states, but they largely operate outside the probabilistic Bayesian assimilation framework, as standalone generative pipelines that forgo the sequential posterior update and rigorous uncertainty quantification guaranteed by Bayes' theorem in traditional filtering. Score-PFF builds on this insight by integrating neural network-learned score functions within the established PFF framework, using $\mathbf{s}_\theta(\mathbf{x})$ as a plug-in replacement for the Gaussian prior gradient while retaining sequential Bayesian structure. This strategy explores the potential of score-based models as enhancements to operational systems, rather than complete replacements.
	
	Numerical experiments on the Lorenz-96 system demonstrate superior accuracy 
	and efficiency compared to both \citeA{hu2021particle} and the Ensemble 
	Adjustment Kalman Filter \cite[EAKF]{Anderson_2001,Anderson_2003}. The remainder of this paper is organized as follows: Section~\ref{sec:method} introduces the methodology; Section~\ref{sec:experiments} presents numerical experiments based on the Lorenz-96 model; and finally, conclusions and future perspectives are provided.

\section{Method}\label{sec:method}

\subsection{Particle Flow Filter Foundation}\label{sec:pff_foundation}

Traditional particle filters suffer from weight degeneracy in high dimensions, necessitating resampling that destroys particle diversity. The PFF circumvents this by transporting particles continuously from prior to posterior through a velocity field, preserving the ensemble size without resampling.

Here, "particles" refer to possible realizations of the system state, i.e., sample points in the state space; "particle flow" describes the continuous evolution trajectory of these sample points in pseudo-time $s \in [0,1]$, which parametrizes the interpolation from prior ($s=0$) to posterior ($s=1$) independently of physical model time. To derive a tractable velocity field, the particle flow is embedded in a reproducing kernel Hilbert space, where the optimal update is obtained by minimizing the Kullback-Leibler divergence between the prior and posterior distributions:
\begin{equation}
	\frac{d}{ds}\mathbf{x}_s = \mathbf{f}_s(\mathbf{x}_s)
	\label{eq:particle_flow}
\end{equation}
where $\mathbf{f}_s(\mathbf{x}_s)$ is the particle flow velocity field, with the specific form derived in \citeA[Appendix~A]{hu2021particle}:
\begin{equation}
	\mathbf{f}_s(\mathbf{x}) = \mathbf{D} \left[ \frac{1}{N_p}\sum_{i=1}^{N_p} \left( \mathbf{K}(\mathbf{x}_s^i, \mathbf{x}) \nabla_{\mathbf{x}_s^i} \log p(\mathbf{x}_s^i|\mathbf{y}) + \nabla_{\mathbf{x}_s^i} \cdot \mathbf{K}(\mathbf{x}_s^i, \mathbf{x}) \right) \right]
	\label{eq:velocity_field}
\end{equation}
In Eq.~\eqref{eq:velocity_field}, $\mathbf{D}$ is a positive definite preconditioning matrix (typically the localized prior covariance $\mathbf{B}_{\text{loc}}$) that scales the update magnitude; $\mathbf{K}$ is the kernel function encoding inter-particle interactions; $\nabla \log p(\mathbf{x}|\mathbf{y})$ is the posterior log-gradient driving particles toward high-probability regions; and the kernel divergence term $\nabla \cdot \mathbf{K}$ generates repulsive forces to maintain particle diversity, where the index $i=1,\ldots,N_p$ denotes the $i$-th particle in the ensemble.

\subsection{Matrix-Valued Kernel Improvement and Its Limitations}
\label{sec:matrix_valued_kernel}

To alleviate marginal collapse, \citeA{hu2021particle} proposed matrix-valued kernel functions that assign dimension-specific bandwidths:
\begin{equation}
	\mathbf{K}(\mathbf{x},\mathbf{z}) = \text{diag}\left(k_1(\mathbf{x},\mathbf{z}), \ldots, k_{n_x}(\mathbf{x},\mathbf{z})\right)
	\label{eq:matrix_valued_kernel}
\end{equation}
where $k_j(\mathbf{x},\mathbf{z}) = \exp\left(-\frac{(x_j-z_j)^2}{2\alpha\sigma_j^2}\right)$, with $\sigma_j$ being the prior standard deviation of the $j$-th dimension and $\alpha$ a global scaling hyperparameter. This anisotropic structure allows independent adaptation to local density variations in each dimension, enabling PFF to scale to systems with $10^5$--$10^6$ state variables.

Despite this architectural advance, existing PFF frameworks remain constrained by Gaussian assumptions in the prior representation. By Bayes' theorem, the posterior log-gradient decomposes as:
\begin{equation}
	\nabla_{\mathbf{x}} \log p(\mathbf{x}|\mathbf{y}) = \underbrace{\nabla_{\mathbf{x}} \log p(\mathbf{y}|\mathbf{x})}_{\text{likelihood gradient}} + \underbrace{\nabla_{\mathbf{x}} \log p(\mathbf{x})}_{\text{prior gradient}}
	\label{eq:posterior_gradient_decomposition}
\end{equation}

For Gaussian observation errors, the likelihood gradient is analytically tractable:
\begin{equation}
	\nabla_{\mathbf{x}} \log p(\mathbf{y}|\mathbf{x}) = \mathbf{H}^\mathsf{T} \mathbf{R}^{-1} (\mathbf{y} - \mathbf{H}(\mathbf{x}))
	\label{eq:likelihood_gradient}
\end{equation}
However, the prior gradient in \citeA{hu2021particle} explicitly assumes a Gaussian prior:
\begin{equation}
	\nabla_{\mathbf{x}} \log p(\mathbf{x}) = -\mathbf{B}^{-1}(\mathbf{x}-\bar{\mathbf{x}})
	\label{eq:gaussian_prior_gradient}
\end{equation}
where $\mathbf{B}$ is the prior covariance matrix (typically the localized covariance $\mathbf{B}_{\text{loc}}$ in practice), and $\bar{\mathbf{x}}$ is the (prior) ensemble mean. This parametric form cannot characterize non-Gaussian structures (multimodal or skewed distributions) inherent in chaotic dynamics. Furthermore, the explicit covariance inversion required by the Gaussian parameterization becomes prohibitive for large $n_x$, even with localization.

\subsection{Score-PFF: Score-Based Particle Flow Filter}
\label{sec:score_pff}

To address the dual limitations of Gaussian prior assumptions and computational bottlenecks in existing PFF, this study proposes the  Score-PFF. The new method retains the matrix-valued kernel framework and Bayesian gradient decomposition of \citeA{hu2021particle}, but replaces the parametric Gaussian prior gradient (Eq.~\eqref{eq:gaussian_prior_gradient}) with a non-parametric score function learned by deep neural networks. This modification eliminates explicit covariance matrix construction and inversion, reducing computational complexity while enabling characterization of non-Gaussian, multimodal prior structures.

\subsubsection{Score Network Training}
\label{sec:score_network_training}

The score network is trained on state samples generated from free runs of the target dynamical model. Given a training sample $\mathbf{x}_0$, noise is injected at multiple scales $\sigma \in \{\sigma_1, \ldots, \sigma_L\}$ to produce perturbed states $\mathbf{x}_\sigma = \mathbf{x}_0 + \sigma \mathbf{z}$ with $\mathbf{z} \sim \mathcal{N}(\mathbf{0}, \mathbf{I})$, where $\sigma$ is sampled from the predefined set of $L$ noise levels. A conditional network $\mathbf{s}_\theta(\mathbf{x}, \sigma)$ learns to predict the score function:
\begin{equation}
	\mathbf{s}_\theta(\mathbf{x}, \sigma) \approx \nabla_{\mathbf{x}} \log p(\mathbf{x})
	\label{eq:score_approximation}
\end{equation}
via denoising score matching (DSM):
\begin{equation}
	\mathcal{L}(\theta) = \mathbb{E}_{\mathbf{x}_0, \sigma, \mathbf{z}} \left[ \left\| \mathbf{s}_\theta(\mathbf{x}_0 + \sigma\mathbf{z}, \sigma) + \frac{\mathbf{z}}{\sigma} \right\|_2^2 \right].
	\label{eq:dsm_loss}
\end{equation}
The multi-scale noise conditioning enables the network to capture both coarse and fine structures of the data distribution.

Network architecture selection depends on the structure of the target dynamical system. For the Lorenz-96 model used in this study, three architectures are evaluated: multilayer perceptrons (MLP) as a generic baseline; Fourier Neural Operators (FNO) that exploit the cyclic symmetry of the latitude-circle discretization through spectral convolutions; and Graph Neural Networks (GNN) that explicitly model local neighbor interactions in the state space. These choices allow assessment of how inductive biases affect data efficiency and assimilation accuracy. Training protocols and hyperparameter settings are detailed in Section~\ref{sec:experiments}.

\subsubsection{Score-PFF Inference}
\label{sec:score_pff_inference}

During online assimilation, the trained score network replaces the Gaussian prior gradient (Eq.~\eqref{eq:gaussian_prior_gradient}). The posterior log-gradient combines the exact likelihood gradient with the neural network-approximated prior gradient:
\begin{equation}
	\nabla_{\mathbf{x}} \log p(\mathbf{x}|\mathbf{y}) = \underbrace{\mathbf{H}^\mathsf{T}\mathbf{R}^{-1}(\mathbf{y}-\mathbf{H}(\mathbf{x}))}_{\text{likelihood gradient (exact)}} + \underbrace{\mathbf{s}_\theta(\mathbf{x},\sigma_{\min})}_{\text{prior gradient (learned)}}
	\label{eq:score_posterior_gradient}
\end{equation}
where $\sigma_{\min}$ is the minimum noise level from training (0.01 in this study). At this scale, the perturbed training distribution $p(\mathbf{x}_0 + \sigma_{\min}\mathbf{z})$ closely approximates the true data distribution $p(\mathbf{x}_0)$, enabling accurate score estimation for clean states encountered during inference.

This modification eliminates the SVD-based covariance inversion required by the Gaussian prior gradient. The prior gradient is obtained via a single network forward pass, avoiding the numerical instabilities and parameter sensitivity of truncated SVD. While the nominal asymptotic complexity remains comparable for localized matrices, the neural network replaces sequential matrix factorization with dense linear algebra that is amenable to GPU acceleration, offering potential for further speedup in GPU-equipped operational settings. Substituting Eq.~\eqref{eq:score_posterior_gradient} into the particle flow equation (Eq.~\eqref{eq:velocity_field}), particles initialized from the prior ensemble are driven toward the posterior through $N_ s$ pseudo-time integration steps. The complete Score-PFF procedure is detailed in \textbf{Algorithm~\ref{alg:score_pff}}.
		
	\begin{algorithm}[t]
		\KwRequire{Prior particle set $\{\mathbf{x}_0^i\}_{i=1}^{N_p}$, observation $\mathbf{y}$, observation operator $\mathbf{H}$, observation error covariance $\mathbf{R}$, trained score network $\mathbf{s}_\theta$, pseudo time step $\Delta s$, number of iterations $N_s$}
		\KwEnsure{Posterior particle set $\{\mathbf{x}_{N_s}^i\}_{i=1}^{N_p}$}
		
		\textbf{Initialization:} Compute prior mean $\mathbf{x}_b = \frac{1}{N_p}\sum_{i=1}^{N_p}\mathbf{x}_0^i$, estimate preconditioning matrix $\mathbf{D}=\mathbf{B}_{\text{loc}}$\;
		
		\For{$s = 1$ \KwTo $N_s$}{
			\For{each particle $i = 1,\ldots,N_p$ \textbf{in parallel}}{
				$\mathbf{g}_{\text{obs}}^i \leftarrow \mathbf{H}^\top\mathbf{R}^{-1}(\mathbf{y}-\mathbf{H}(\mathbf{x}_{s-1}^i))$\Comment*[r]{Likelihood gradient}
				$\mathbf{g}_{\text{prior}}^i \leftarrow \mathbf{s}_\theta(\mathbf{x}_{s-1}^i,\sigma_{\min})$\Comment*[r]{Prior gradient via score network}
				$\nabla\log p^i \leftarrow \mathbf{g}_{\text{obs}}^i + \mathbf{g}_{\text{prior}}^i$\Comment*[r]{Total posterior gradient}
			}
			\For{each particle $i = 1,\ldots,N_p$ \textbf{in parallel}}{
				Compute matrix-valued kernel $\mathbf{K}(\mathbf{x}_{s-1}^i,\mathbf{x}_{s-1}^j)$ for $j=1,\ldots,N_p$ via Eq.~\eqref{eq:matrix_valued_kernel}\;
				$\mathbf{f}^i \leftarrow \mathbf{D}\left[\frac{1}{N_p}\sum_{j=1}^{N_p}\left(\mathbf{K}(\mathbf{x}_{s-1}^j,\mathbf{x}_{s-1}^i)\nabla\log p^j + \nabla_{\mathbf{x}_{s-1}^j}\cdot\mathbf{K}(\mathbf{x}_{s-1}^j,\mathbf{x}_{s-1}^i)\right)\right]$\Comment*[r]{Particle flow velocity}
				$\mathbf{x}_s^i \leftarrow \mathbf{x}_{s-1}^i + \Delta s \cdot \mathbf{f}^i$\Comment*[r]{Particle update}
			}
		}
		\Return{$\{\mathbf{x}_{N_s}^i\}_{i=1}^{N_p}$}\;
		\caption{Score-based Particle Flow Filter (Score-PFF)}
		\label{alg:score_pff}
	\end{algorithm}
		
	\section{Numerical Experiments and Results}\label{sec:experiments}
	
	\subsection{Experimental Design}\label{sec:experimental_design}
	
	\subsubsection{Dynamical Model}
	
	Numerical experiments employ the Lorenz-96 model as the testbed. Proposed by \citeA{Lorenz_1996}, this model constitutes a multiscale chaotic system representing a low-order approximation of zonal atmospheric dynamics. Its governing equations are:
	\begin{equation}\label{eq:l96}
		\frac{\mathrm{d} x_{k}}{\mathrm{d} t} = -x_{k-1}\left(x_{k-2}-x_{k+1}\right) - x_{k} + F, \quad k=1,\ldots,K
	\end{equation}
	where $x_k$ denotes the state variable of the $k$-th sector after dividing the latitude circle into $K$ equal sectors, with periodic boundary conditions ($x_{0}=x_{K}$, $x_{-1}=x_{K-1}$, $x_{K+1}=x_{1}$). The three terms on the right-hand side represent advective transport, damping dissipation, and external forcing, respectively. When the forcing parameter $F=8$, the system exhibits chaotic behavior with an error doubling time of approximately 0.4 time units (TU), corresponding to roughly 5 days of atmospheric predictability \cite{Lorenz_Emanuel_1998}. Model integration employs the fourth-order Runge-Kutta scheme with a time step $\Delta t = 0.01$ TU.
	
	\subsubsection{Observations and Observation Operators}
	
	Following the experimental design of \citeA{hu2021particle}, observations are constructed as follows. First, starting from random initial conditions, the model is integrated for 1000 steps (10 TU) to eliminate transient effects and allow the system to reach statistical equilibrium; subsequent integration serves as the true trajectory $\mathbf{x}^t$. Observations are then sampled from this trajectory. This study adopts a spatially sparse observation strategy: one observation is placed every 4 grid points, yielding an observation coverage of 25\%. For the 40-dimensional system, the number of observed variables is 10; for the 1000-dimensional system, it is 250. The observation interval is 20 model time steps (0.2 TU), corresponding to an observation frequency of approximately 1 day.
	
	Under the linear observation operator, observations are generated by superimposing Gaussian white noise onto the true values:
	$$
	y_k = \mathcal{H}(x_k) + \varepsilon_k, \quad \varepsilon_k \sim \mathcal{N}(0, \sigma_o^2)
	$$
	where $\sigma_o$ is the observation error standard deviation. Beyond the standard linear operator $\mathcal{H}(x)=x$, three nonlinear observation operators are designed to test performance under strongly non-Gaussian conditions:
	\begin{itemize}
		\item \textbf{Absolute-value observation}: $\mathcal{H}(x) = |x|$ 
		\item \textbf{Squared observation}: $\mathcal{H}(x) = x^2$ 
		\item \textbf{Exponential observation}: $\mathcal{H}(x) = \exp(x/6)$ 
	\end{itemize}
	
	\subsubsection{Assimilation Configuration}
	
	This study designs two complementary experiments to validate distinct aspects of the proposed method. The 40-dimensional experiment ($K=40$, each grid point corresponding to a $9^\circ$ longitude band) focuses on verifying Score-PFF's capability to characterize non-Gaussian distributions. The 1000-dimensional experiment ($K=1000$ ) tests the practical computational benefits of eliminating SVD-based covariance inversion in high-dimensional sparse observation scenarios, where this operation becomes a severe bottleneck for Gaussian PFF.
	
	The 40-dimensional experiments employ 30 ensemble members to ensure robust performance across all methods under nonlinear observation conditions. The 1000-dimensional experiments employ 20 ensemble members, consistent with \citeA{hu2021particle} for direct comparability. All methods utilize the Gaspari-Cohn localization function, with localization radius $r_{\mathrm{loc}}=4$ grid points for the 40-dimensional system and $r_{\mathrm{loc}}=12$ grid points for the 1000-dimensional system. 
	Specifically, PFF employs localized $\mathbf{B}$ for three purposes: (i) prior gradient computation via $-\mathbf{B}_{\text{loc}}^{-1}(\mathbf{x}-\bar{\mathbf{x}})$, (ii) preconditioning matrix $\mathbf{D}=\mathbf{B}_{\text{loc}}$ in Eq.~\eqref{eq:velocity_field}, and (iii) kernel bandwidth $\sigma_j^2=\text{diag}(\mathbf{B}_{\text{loc}})_j$ in Eq.~\eqref{eq:matrix_valued_kernel}. Score-PFF retains only (ii)--(iii), replacing (i) with the neural network-learned score $\mathbf{s}_\theta(\mathbf{x})$. 
	
	EAKF requires a multiplicative inflation factor of 1.25 applied to the localized background error covariance matrix to maintain ensemble spread; in contrast, both PFF variants operate without inflation, as particle flow dynamics inherently preserve ensemble diversity through kernel-mediated particle interactions.

\subsection{Results for the 40-Dimensional System}

Based on the experimental configuration in Section~\ref{sec:experimental_design}, this section presents a detailed analysis of the 40-dimensional assimilation results. We first examine the score network's capability to characterize non-Gaussian attractor structure through a pure prior adjustment experiment without observations. We then proceed to data assimilation experiments under linear and nonlinear observation operators.

\subsubsection{Pure Prior Adjustment Experiment}\label{sec:prior_adjustment}

The score network is trained using DSM. Training data are generated from free runs of the 40-dimensional Lorenz-96 model, comprising 500 trajectories with 1 million state samples. Each trajectory consists of 2000 steps collected after a 1000-step spin-up. The network architecture is a four-layer MLP with hidden layer dimensions $[128, 256, 256, 128]$, which is independent of the state dimension $n_x$. The network takes the noise-perturbed state $\mathbf{x}_\sigma = \mathbf{x}_0 + \sigma\mathbf{z}$ and the noise level $\sigma$ as inputs to learn the score function $\mathbf{s}_\theta(\mathbf{x}_\sigma, \sigma) \approx \nabla_{\mathbf{x}} \log p(\mathbf{x})$, with the target $-\mathbf{z}/\sigma$ derived from the known noise realization $\mathbf{z} \sim \mathcal{N}(\mathbf{0}, \mathbf{I})$. Training employs the Adam optimizer with an initial learning rate of $10^{-3}$ and a cosine annealing schedule, a batch size of 256, and early stopping. The model converges after 39 epochs.

\begin{figure}
			\noindent\includegraphics[width=\textwidth]{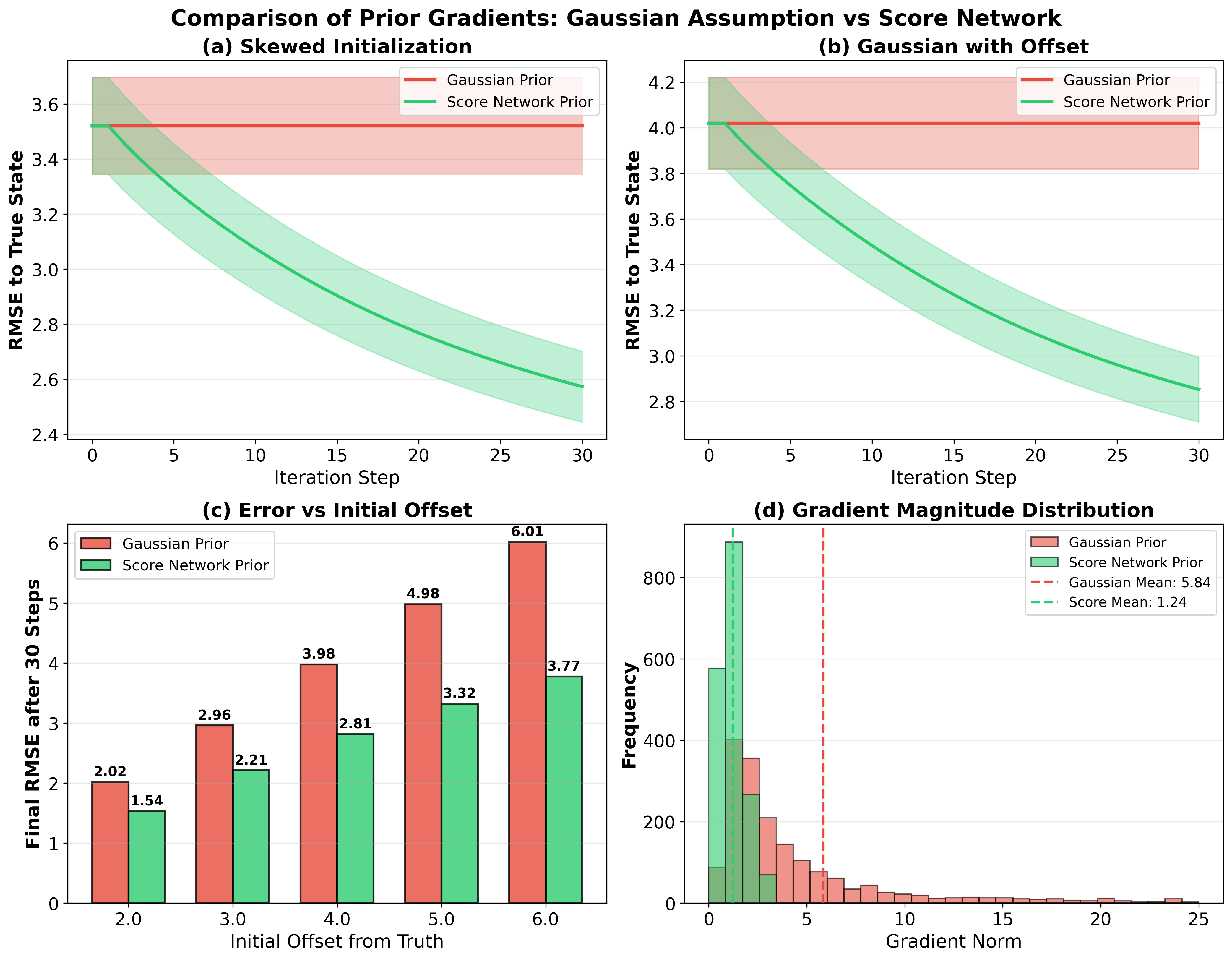}
			\caption{Pure prior adjustment experiment comparing Gaussian and neural network priors on the 40-dimensional Lorenz-96 system. (a) Skewed initialization: all particles initialized on one side of the truth; (b) Offset Gaussian initialization: Gaussian distribution with center offset by 4.0 from the truth; (c) Improvement rate as a function of initial offset distance (2.0 to 6.0); (d) Gradient norm distribution for the neural network prior and Gaussian prior.}\label{fig:prior_infer}
		\end{figure}

To verify the score network's capability to characterize non-Gaussian attractor features independently of observation information, this study designs a pure prior adjustment experiment using the trained network with 30 particles and 30 iterations without observations. The results are shown in Figure~\ref{fig:prior_infer}. Two initializations are tested: (i) skewed, where all particles are placed on one side of the truth, and (ii) offset Gaussian, where the center is displaced from the truth.

For the skewed initialization (Figure~\ref{fig:prior_infer}a), the Gaussian prior fails to reduce the error, which remains at 3.52, as its gradient points toward the sample mean. In contrast, the neural network prior reduces the error to 2.57, representing a 26.9\% improvement. For the offset Gaussian initialization with an offset distance of 4.0 (Figure~\ref{fig:prior_infer}b), the Gaussian prior again fails with an error of 4.02, whereas the network prior achieves an error of 2.85, a 29.0\% improvement. This advantage scales with the offset distance (Figure~\ref{fig:prior_infer}c): the improvement is 24\% at an offset of 2.0 and rises to 37\% at an offset of 6.0.

Gradient analysis (Figure~\ref{fig:prior_infer}d) reveals the underlying mechanism. The score network prior produces gradients concentrated near zero (mean magnitude 1.24), computed by neural network forward pass $\mathbf{s}_\theta(\mathbf{x})$. The Gaussian prior yields broadly dispersed gradients (mean magnitude 5.84), computed analytically as $-\mathbf{B}^{-1}(\mathbf{x}-\bar{\mathbf{x}})$. This disparity demonstrates that the score network learns the true climatological distribution, providing precise gradient directions toward the attractor, whereas the Gaussian prior merely pulls particles toward the sample mean.
		
\subsubsection{Assimilation Results under Linear Observations}\label{sec:experiment_linear}

Building on the validation of the score network's non-Gaussian characterization capability, we proceed to data assimilation experiments with observations. For fair comparison, all experiments employ identical Gaussian-perturbed initial ensembles $\mathbf{x}_0^{(i)} = \mathbf{x}_{\text{true}} + \boldsymbol{\epsilon}_i$, where $\boldsymbol{\epsilon}_i \sim \mathcal{N}(\mathbf{0}, 2\mathbf{I})$, consistent with \citeA{hu2021particle} to ensure flow-dependent initial covariance structure. Particle flow parameters are uniformly set: maximum pseudo-time steps 50, initial step length 0.05, adaptive shrinkage factor 1.5, stopping threshold 5\%, and kernel bandwidth $\alpha=1/N_p$. To enhance numerical stability, we introduce several algorithmic modifications. Gradient clipping limits the $L_2$ norm of gradients to 50. Adaptive step-size scales the step size by 2/3 when the $L_2$ norm exceeds 50. SVD truncated inversion removes singular values below $10^{-5}$ relative to the largest. State constraints bound variables to $[-20,20]$. Each configuration comprises 20 independent replicate trials. Performance is evaluated by time-averaged RMSE, with paired $t$-tests at $\alpha=0.05$ and Cohen's $d$ effect sizes for statistical validation \cite{cohen1992power}. Small, medium, and large effects are defined as $|d|<0.2$, $0.5\leq|d|<0.8$, and $|d|\geq0.8$, respectively.

\begin{figure}
	\noindent\includegraphics[width=\textwidth]{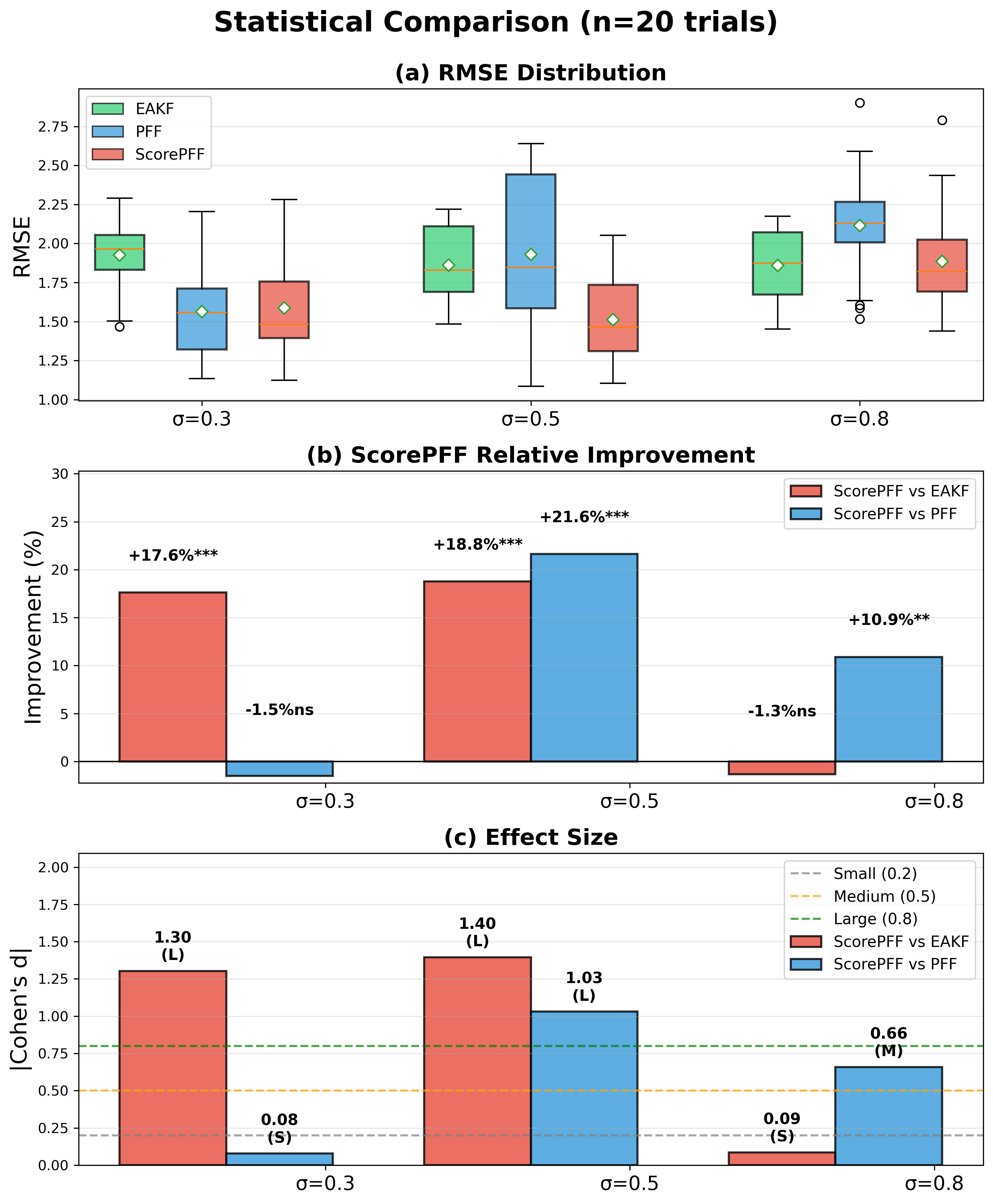}
	\caption{Statistical comparison of EAKF, PFF, and Score-PFF on the 40-dimensional Lorenz-96 system with 30 ensemble members over 20 independent trials. (a) RMSE distribution boxplots under three observation error levels; (b) Relative improvement percentage of Score-PFF over EAKF and PFF, with significance levels annotated (*$p < 0.05$, **$p < 0.01$, ***$p < 0.001$; ns = not significant); (c) Effect sizes for Score-PFF comparisons, with reference lines marking small, medium, and large effects.}\label{fig:stats}
\end{figure}

Figure~\ref{fig:stats} presents statistical comparison across three observation error levels on the 40-dimensional Lorenz-96 system. The error levels are $\sigma = 0.3$, $0.5$, and $0.8$, with 20 trials and 10 assimilation cycles for each configuration.
At $\sigma = 0.3$, corresponding to strong observations, PFF and Score-PFF achieve lower median RMSE than EAKF with overlapping distributions, as shown in Figure~\ref{fig:stats}a. The strong likelihood constraints at this level reduce the influence of prior gradient quality. At $\sigma = 0.5$, corresponding to moderate observations, PFF exhibits elongated boxes with higher median, indicating numerical instability. In contrast, Score-PFF maintains a tight distribution with the lowest median, demonstrating that accurate non-Gaussian prior gradient estimation becomes critical when likelihood constraints are moderate. At $\sigma = 0.8$, corresponding to weak observations, all three methods converge as observation information becomes insufficient to distinguish prior quality.

Paired $t$-tests, shown in Figure~\ref{fig:stats}b, indicate that Score-PFF significantly outperforms EAKF at $\sigma = 0.3$ and $\sigma = 0.5$ with $p < 0.001$, and PFF at $\sigma = 0.5$ with $p = 0.018$. Effect sizes, presented in Figure~\ref{fig:stats}c, confirm large advantages at $\sigma = 0.5$, with $d = 1.03$ versus PFF and $d = 1.40$ versus EAKF. 
These results demonstrate that Score-PFF's neural network-learned prior gradients provide decisive advantages when observation constraints are moderate, enabling effective guidance of particles toward high posterior probability regions that Gaussian priors cannot capture.

\begin{figure}
	\noindent\includegraphics[width=\textwidth]{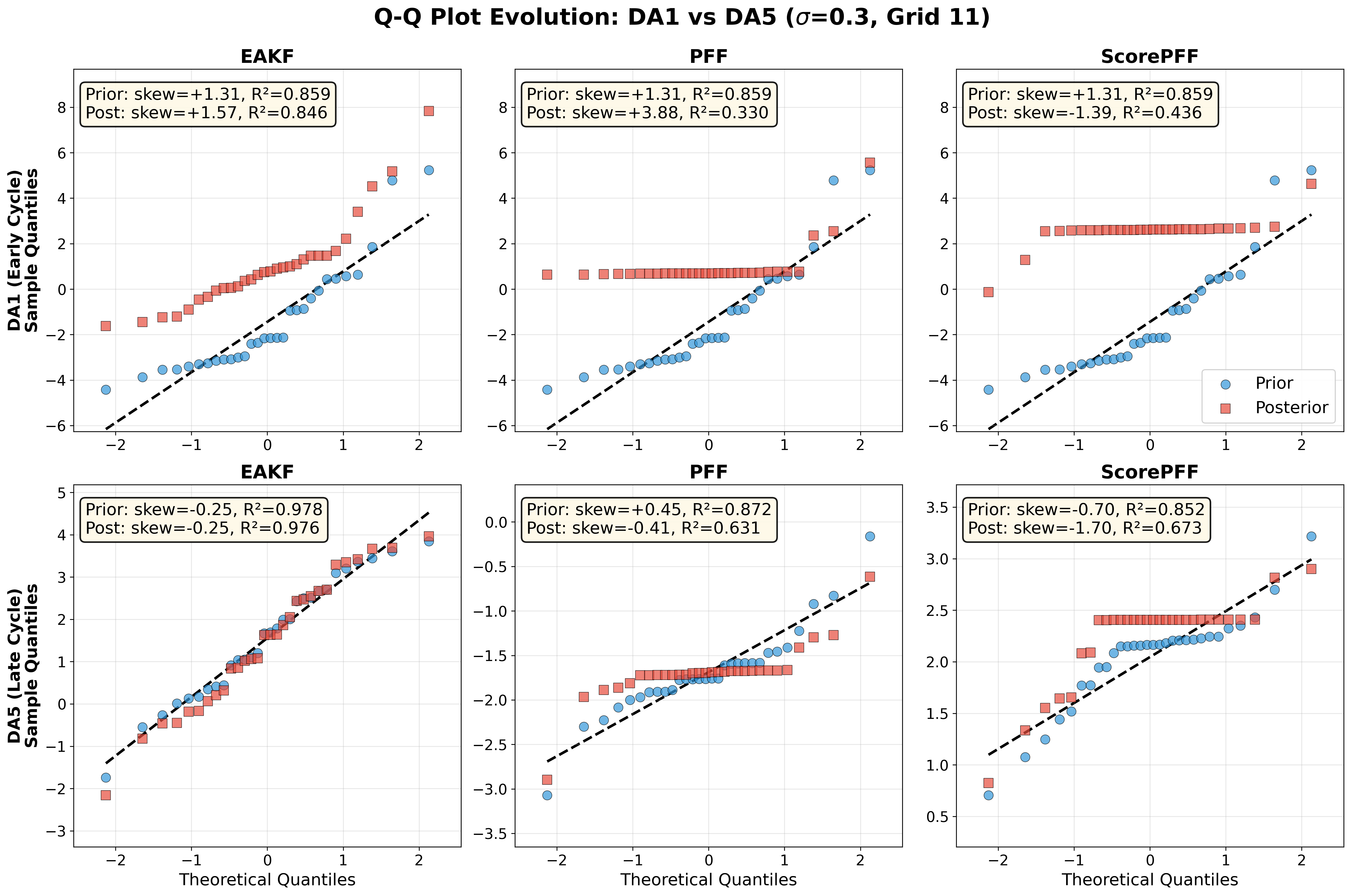}
	\caption{Evolution of Q-Q plots for three assimilation methods in the Lorenz-96 system, with observation error $\sigma=0.3$. Top row shows state distributions at assimilation cycle 1, bottom row at assimilation cycle 5.}\label{fig:qq}
\end{figure}

To diagnose the source of PFF-class advantages over EAKF, we examine ensemble distribution evolution using quantile-quantile (Q-Q) plots. Departures from the diagonal reference line indicate non-Gaussianity, and $R^2$ quantifies the degree of Gaussianity, with values near 1 indicating Gaussian structure and lower values indicating non-Gaussian structure \cite{wilk1968probability}. Figure~\ref{fig:qq} presents Q-Q plots for DA1 and DA5 at grid point 11 with observation error $\sigma = 0.3$. At DA1, all methods share identical non-Gaussian priors, exhibiting a marked S-shaped curvature with skewness of approximately 1.31. By DA5, EAKF Gaussianizes the ensemble (skewness near $-0.25$, $R^2 \approx 0.98$), while PFF and Score-PFF retain non-Gaussian structure (absolute skewness exceeding 0.7, $R^2 \approx 0.86$), demonstrating that particle flow methods preserve the L96 attractor's non-Gaussianity whereas EAKF's linear updates progressively erase it.

\subsubsection{Assimilation Results under Nonlinear Observations}\label{sec:experiment_nonlinear}

This experiment employs three nonlinear observation operators to further examine method performance: absolute-value $\mathcal{H}(x)=|x|$, squared $\mathcal{H}(x)=x^2$, and exponential $\mathcal{H}(x)=\exp(x/6)$. All three methods conduct assimilation experiments with stability measures as indicated in Section~\ref{sec:experiment_linear}.

Notably, the squared observation operator poses inherent challenges near $|x| \approx 10$. The gradient $d\mathcal{H}/dx = 2x$ reaches $\pm 20$, and the double-valued mapping $x = \pm\sqrt{y}$ introduces sign ambiguity, easily causing particle trajectory divergence. Targeted enhancement strategies are therefore implemented for this operator. Gradient clipping is tightened to limit the $L_2$ norm to 10, versus the default limit of 50. Particle flow iterations are extended to 100 steps, versus the default 50, with more conservative adaptive step-size using an initial length of 0.005 halved after 20 steps, versus the default 0.05. State constraints are tightened to bound variables to $[-15, 15]$, versus the default $[-20, 20]$. Additionally, prior score update frequency is increased to every 2 steps in Score-PFF.

\begin{figure}
	\noindent\includegraphics[width=\textwidth]{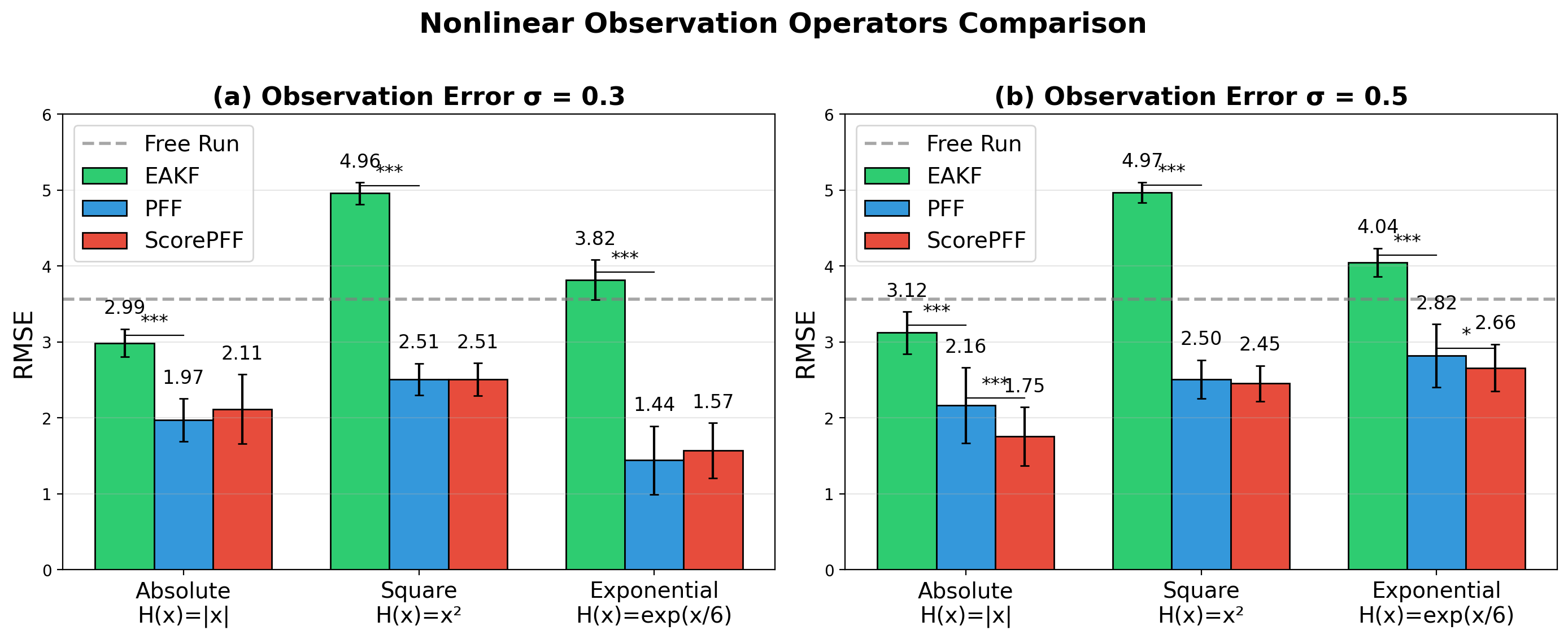}
	\caption{RMSE comparison under nonlinear observations (absolute-value, squared, exponential) for EAKF, PFF, and Score-PFF. Based on 20 replicate experiments per configuration. Statistical significance (paired $t$-test) is annotated above comparison bars: *** $p<0.001$, ** $p<0.01$, * $p<0.05$. Non-significant comparisons ($p\geq0.05$) are omitted from the figure for clarity. (a) $\sigma=0.3$; (b) $\sigma=0.5$.}
	\label{fig:nonlinear}
\end{figure}

To verify method robustness under nonlinear observation conditions, each configuration comprises 20 independent replicate trials, with significance assessed via paired $t$-tests. Assimilation results in Figure~\ref{fig:nonlinear} demonstrate that both PFF variants significantly outperform EAKF across all six experimental conditions, validating the fundamental advantage of particle flow methods in handling strongly non-Gaussian posteriors.

In Figure~\ref{fig:nonlinear}a, exponential observations show the largest improvement of PFF over EAKF at approximately 60\%. Squared observations, which produce a bimodal posterior, follow with approximately 50\% improvement. Absolute-value observations show relatively smaller improvement at approximately 34\%. Notably, EAKF errors under squared and exponential observations even exceed the no-assimilation baseline, indicating severe failure of Gaussian assumptions in these scenarios. At $\sigma = 0.3$, no significant difference exists between the two PFF variants.

However, in Figure~\ref{fig:nonlinear}b with $\sigma = 0.5$, Score-PFF marginally outperforms PFF for absolute-value observations and shows marginal significance for exponential observations. Yet no significant difference remains for squared observations, possibly due to the aggressive stability measures employed by both methods in this challenging scenario. This comparison suggests that when observation error is moderately larger, neural network priors slightly enhance PFF assimilation performance under nonlinear observations, though the difference is less pronounced than in linear scenarios.

		\subsubsection{Impact of Training Data Volume and Model Architecture}\label{sec:experiment_architecture}
		
The above experiments demonstrate that Score-PFF significantly improves assimilation accuracy through neural network-learned non-Gaussian priors, yet its performance depends on score network training quality. The preceding results were obtained with 1 million training samples, whereas in practical Earth system applications, high-fidelity model runs are computationally expensive, making such large-scale datasets difficult to generate. Therefore, we investigate the impact of network architecture design on data efficiency under limited training data.

\begin{figure}
	\noindent\includegraphics[width=\textwidth]{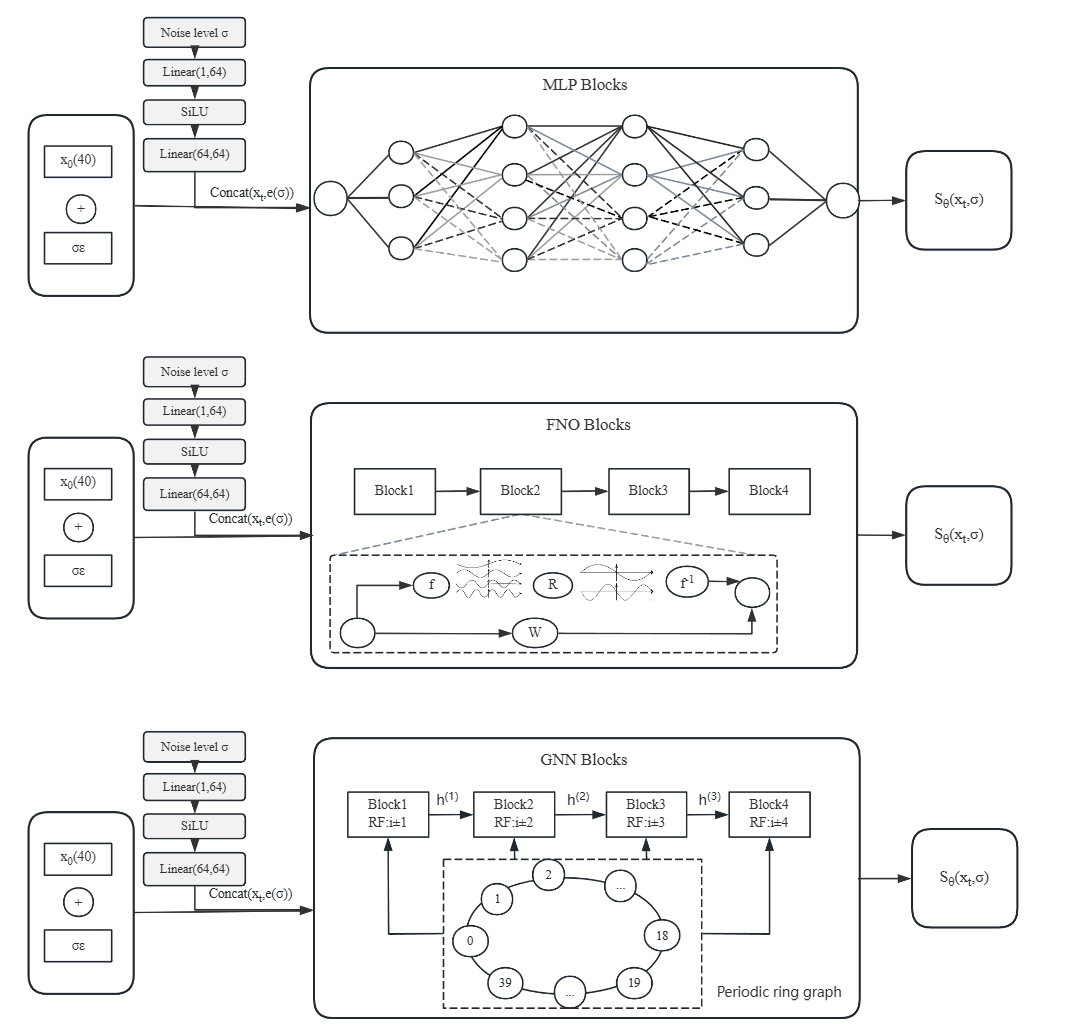}
	\caption{Score network architectures for Score-PFF. (a) MLP: fully connected layers [128, 256, 256, 128]. (b) FNO: spectral convolution with 32 Fourier modes and width 128, leveraging cyclic symmetry. (c) GNN: 4-layer graph convolution with hidden dimension 128, encoding local grid interactions. Inputs are noise-perturbed state $\mathbf{x}_\sigma$ and noise level $\sigma$; output is the score estimate $\mathbf{s}_\theta(\mathbf{x}_\sigma, \sigma)$.}
	\label{fig:diagram}
\end{figure}

\begin{figure}
			\noindent\includegraphics[width=\textwidth]{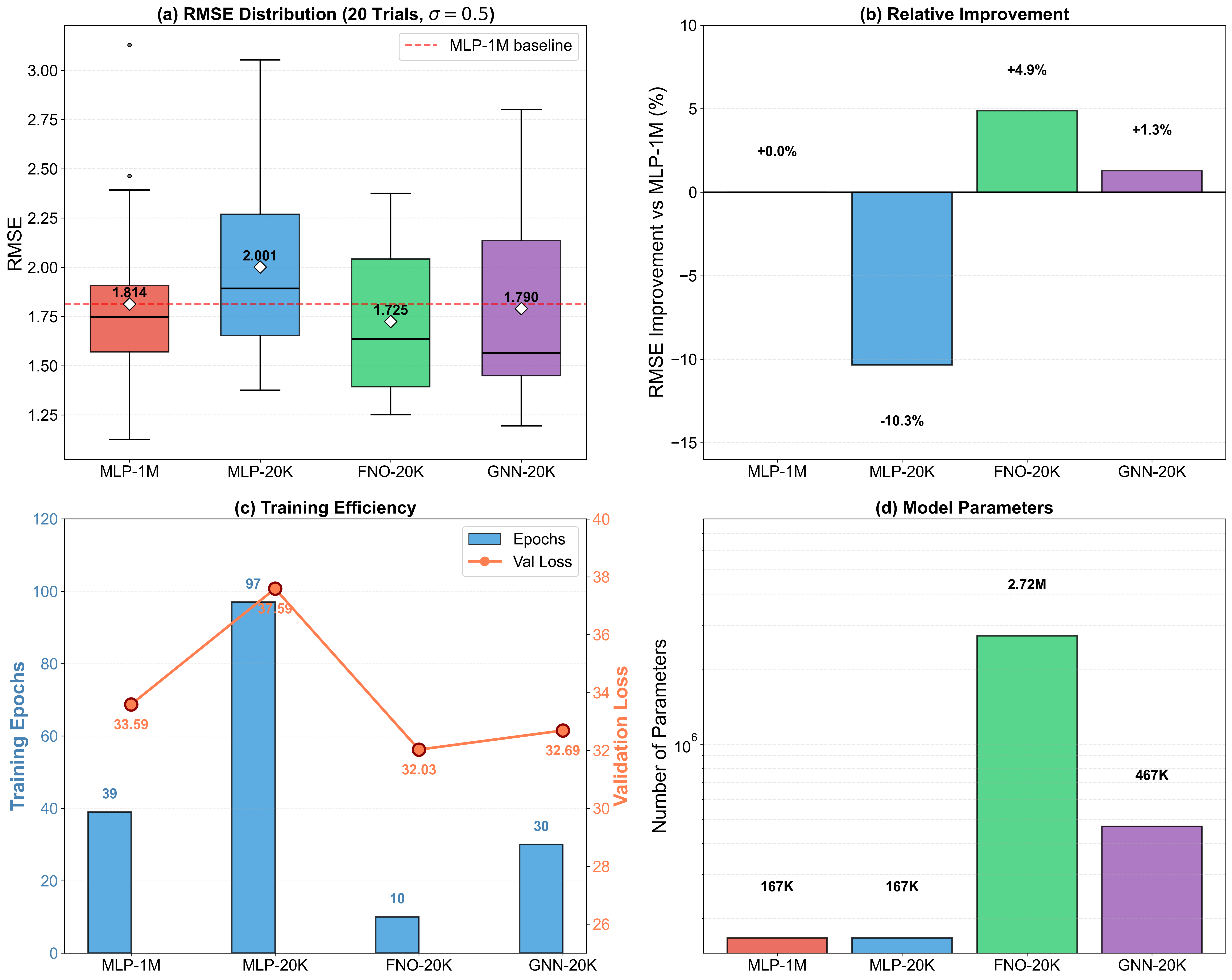}
			\caption{ Comparison of neural network architectures for score-based particle flow filtering on the 40-dimensional Lorenz-96 system with linear observations at $\sigma=0.5$ . (a) RMSE distribution over 20 trials. The dashed line marks the MLP\_1M baseline. (b) Relative improvement over the MLP\_1M baseline. (c) Training efficiency: epochs to convergence and final validation loss. (d) Model parameter count.}
			\label{fig:architecture}
		\end{figure}

Four configurations are compared. The MLP trained with 1 million samples (500 trajectories of length 2000) serves as the baseline. MLP, FNO, and GNN are subsequently trained with 20,000 samples (10 trajectories of length 2000). The FNO uses 32 Fourier modes and width 128, while the GNN employs 4-layer graph convolution with hidden dimension 128. These architectures exploit the cyclic symmetry and local graph structure of L96, respectively (Figure~\ref{fig:diagram}). Under early stopping, FNO converges in 10 epochs, GNN in approximately 30 epochs, while MLP\_20k requires 97 epochs with higher validation loss (Figure~\ref{fig:architecture}c).

Experimental results based on 20 independent trials (Figure~\ref{fig:architecture}) show that FNO\_20k achieves the lowest median RMSE (approximately 1.73) with a concentrated distribution, followed by GNN\_20k (approximately 1.79). MLP\_20k performs worst with notable outliers. Compared to the MLP\_1M baseline, MLP\_20k suffers 10.3\% RMSE degradation due to insufficient data, whereas FNO\_20k and GNN\_20k maintain favorable accuracy, achieving improvements of 4.9\% and 1.3\%, respectively. However, FNO has 2.72M parameters (16 times the MLP parameter count), while GNN has 467K parameters (2.8 times the MLP parameter count), suggesting FNO may become a bottleneck for high-dimensional scaling.

These results demonstrate that domain-specific architectural choices critically affect data efficiency. FNO\_20k achieves optimal performance and training efficiency through frequency-domain modeling, with 10-epoch convergence and the lowest validation loss (32.03). This highlights the advantage of physics-informed architectures in small-sample scenarios. GNN\_20k offers a favorable balance between accuracy and model complexity. MLP\_20k suffers from low data efficiency due to lacking structural priors. For high-dimensional scaling, GNN or MLP\_1M may represent more robust choices.
		
\subsection{Results for the 1000-Dimensional System}

For the 1000-dimensional Lorenz-96 system, this study employs GNN as the score model architecture to verify data efficiency under training data constraints. Unlike the 1 million samples used in the 40-dimensional experiments, this experiment uses only 100,000 samples (50 trajectories of length 2000 collected after a 2000-step spin-up). This reduced sample size simulates the realistic constraint in Earth system applications where high-fidelity model runs are computationally expensive. GNN explicitly models local neighborhood interactions in Lorenz-96, capturing adjacent grid point correlations under periodic boundary conditions through circular padding one-dimensional convolution layers.

The network configuration comprises a 4-layer encoder-decoder architecture with hidden dimension 128 and approximately 470K total parameters. Residual connections and layer normalization stabilize deep training. We employ the same DSM loss function as the 40-dimensional experiments, with 20 noise levels logarithmically sampled in $\sigma \in [0.01, 1.0]$. The batch size is 64, and the Adam optimizer is used with learning rate $10^{-3}$ and weight decay $10^{-5}$. GNN converges in approximately 30 epochs under early stopping.

This study compares PFF and Score-PFF under linear observation operator $\mathcal{H}(x) = x$ with 25\% observation coverage and observation error standard deviation 0.5. Both methods employ identical assimilation configurations: ensemble size 20, assimilation interval of 20 time steps for 15 assimilations total, maximum pseudo-time iterations 20, initial step length $\epsilon = 0.02$, convergence threshold set to 0.1 times the gradient norm, and Gaspari-Cohn localization radius 12. Neither method uses covariance inflation. The sole difference lies in prior gradient computation: PFF uses the Gaussian assumption $-\mathbf{B}_{\text{loc}}^{-1}(\mathbf{x}-\bar{\mathbf{x}})$, while Score-PFF employs the neural network-learned score function $\mathbf{s}_\theta(\mathbf{x})$.

\begin{figure}
	\noindent\includegraphics[width=\textwidth]{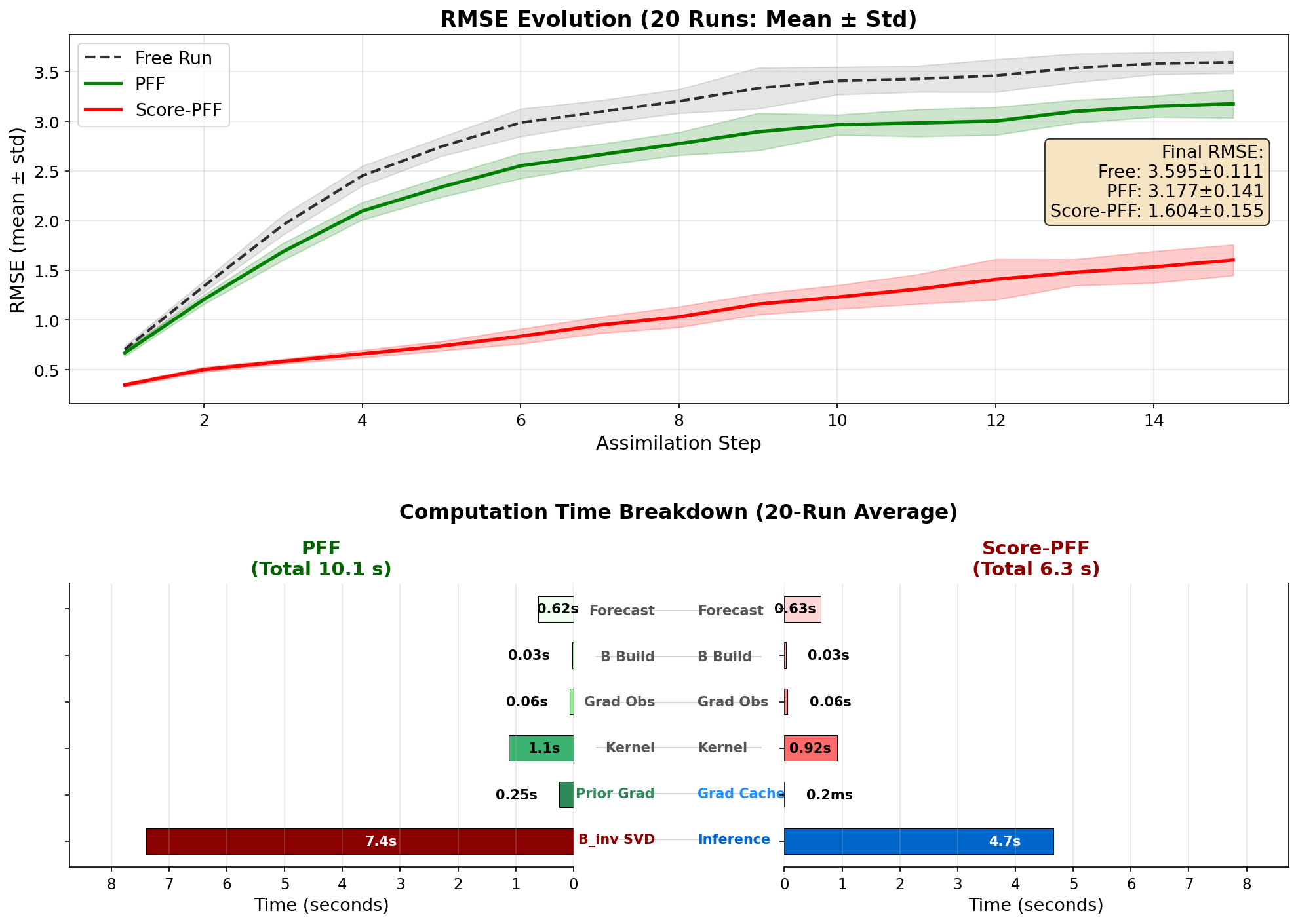}
	\caption{Performance comparison on 1000D Lorenz-96. (a) RMSE over 15 assimilation cycles. (b) Computational time breakdown. }
	\label{fig:1000d}
\end{figure}

Based on 20 independent runs (Figure~\ref{fig:1000d}), Score-PFF significantly outperforms PFF in both assimilation accuracy and computational efficiency. Score-PFF achieves a final RMSE of $1.604 \pm 0.155$, representing a 49.5\% reduction compared to PFF's $3.177 \pm 0.141$, and a 55.4\% reduction compared to the no-assimilation Free Run ($3.595 \pm 0.111$). The difference is statistically significant.

In total computational time, Score-PFF requires 6.3 seconds compared to 10.1 seconds for PFF, yielding a speedup factor of 1.6 (Figure~\ref{fig:1000d}b). This efficiency gain stems from replacing the SVD-based covariance inversion in PFF (7.4s, 73.3\% of total time) with neural network inference (4.7s). The elimination of SVD reduces the dominant computational bottleneck, enabling practical deployment in high-dimensional operational systems without sacrificing accuracy.

Beyond the raw speedup, Score-PFF offers several practical advantages over localized PFF. First, the Gaussian prior gradient $-\mathbf{B}_{\text{loc}}^{-1}(\mathbf{x}-\bar{\mathbf{x}})$ requires careful tuning of the localization radius $r_{\text{loc}}$: too small suppresses true long-range correlations, while too large introduces spurious correlations and degrades filter performance. This parameter sensitivity demands extensive trial-and-error calibration for each new system. Moreover, without localization, the full covariance inversion incurs $\mathcal{O}(n_x^3)$ complexity, which is computationally prohibitive for operational systems with $n_x \sim 10^5$--$10^6$. Second, even with localization, the truncated SVD introduces numerical instabilities sensitive to the effective rank $r$, whose optimal value varies with system dimension and correlation structure. Third, the explicit construction and storage of $\mathbf{B}_{\text{loc}}$ consumes $\mathcal{O}(n_x^2)$ memory regardless of localization, limiting scalability. In contrast, Score-PFF bypasses these issues entirely: the score network requires no localization parameter tuning, no rank truncation, and no dense covariance storage. The prior gradient is obtained from a single forward pass with fixed architecture, offering predictable and robust performance across different problem settings.

The 1000-dimensional experiments were conducted on an Intel Xeon Platinum 8276 processor with 256GB system memory. The software environment was Python 3.9 and NumPy 1.24. The neural network forward pass is amenable to GPU acceleration, which may yield further speedup for larger network architectures or batch processing scenarios.

\section{Conclusion}

This study presents the Score-PFF, a novel data assimilation framework that addresses two fundamental limitations of existing particle flow methods: (1) restrictive Gaussian prior assumptions that fail to capture non-Gaussian attractor structures in chaotic systems, and (2) prohibitive computational bottlenecks arising from SVD-based covariance inversion.

By replacing the parametric Gaussian prior gradient $-\mathbf{B}_{\text{loc}}^{-1}(\mathbf{x}-\bar{\mathbf{x}})$ with a neural network-learned score function $\mathbf{s}_\theta(\mathbf{x}, \sigma_{\min})$ trained via denoising score matching, Score-PFF achieves three key advances:

First, the score network captures multimodal, skewed, and complex prior distributions that Gaussian parameterizations cannot. Pure prior adjustment experiments demonstrate that the score network provides correct gradient directions toward the climatological attractor regardless of initial ensemble configuration, achieving 26.9\%--29.0\% error reduction over Gaussian priors when ensembles are offset from the attractor. Under linear observations with moderate error ($\sigma=0.5$), Score-PFF significantly outperforms both Gaussian-based PFF and EAKF, with Cohen's $d$ effect sizes of 1.03 and 1.40 respectively. Quantile-quantile analysis confirms that Score-PFF preserves non-Gaussian structure over successive assimilation cycles, whereas EAKF progressively Gaussianizes the ensemble.

Second, under nonlinear observation operators (absolute-value, squared, exponential), Score-PFF maintains robust performance with up to 60\% RMSE reduction relative to EAKF in strongly non-Gaussian regimes. The neural network prior eliminates the sign ambiguity and gradient explosion issues that challenge Gaussian approximations, enabling stable assimilation where EAKF fails entirely.

Third, the elimination of SVD-based covariance inversion reduces the dominant computational bottleneck of Gaussian PFF. On the 1000-dimensional Lorenz-96 system, Score-PFF achieves a speedup factor of 1.6 (6.3s versus 10.1s per 20-run experiment) by replacing SVD inversion (7.4s, 73.3\% of PFF time) with neural network inference (4.7s). Beyond raw speedup, Score-PFF avoids the parameter sensitivity of localization radius tuning, the numerical instabilities of truncated SVD rank selection, and the $\mathcal{O}(n_x^2)$ memory requirement of dense covariance storage. The neural network forward pass is amenable to GPU acceleration, offering potential for further speedup in accelerated computing environments.

Domain-specific architectures critically affect data efficiency: Fourier Neural Operators achieve optimal accuracy with minimal training data by exploiting cyclic symmetry, while Graph Neural Networks offer favorable accuracy--complexity trade-offs for high-dimensional scaling. These results demonstrate that incorporating physics-informed inductive biases is essential for operational deployment under realistic training constraints.

 Several challenges remain for operational deployment. First, Score-PFF requires offline training on model simulations, which may be computationally expensive for high-fidelity Earth system models. The 1000-dimensional experiments used 100,000 training samples (50 trajectories), but operational weather and ocean models may require substantially more data to capture multi-scale variability. Second, the score network may suffer from generalization degradation when system parameters (e.g., forcing $F$ in Lorenz-96) deviate significantly from training conditions. Online adaptation strategies using streaming assimilation data warrant investigation. Third, the current stability safeguards---gradient clipping, adaptive step-size, and state constraints---are empirically tuned for Lorenz-96. Automated or learned stability policies would enhance robustness across different dynamical systems. Fourth, while the GNN architecture scales favorably in our experiments, the $\mathcal{O}(L n_x^2)$ complexity of fully connected layers in MLP-based score networks may become prohibitive for million-dimensional operational systems; sparse or factorized architectures require further exploration. Finally, the current experiments employ perfect-model assumptions; real-world applications must account for model error, which may degrade score network performance if training and operational model configurations diverge.

Future work will address these limitations through: (1) online adaptation of score networks using cycling assimilation data; (2) extension to coupled ocean-atmosphere systems with $10^6$--$10^7$ state variables; (3) integration with differentiable numerical models for end-to-end learning of physics-informed priors; and (4) validation in real-world satellite altimetry and sea surface temperature assimilation systems.

In conclusion, Score-PFF establishes a new paradigm for high-dimensional non-Gaussian data assimilation by synergistically combining the mathematical rigor of particle flow methods with the representational power of score-based generative models. The demonstrated accuracy gains in non-Gaussian regimes---together with practical computational efficiency and elimination of sensitive tuning parameters---position this framework as a viable pathway toward next-generation operational assimilation systems that can fully exploit the non-Gaussian information content of geophysical flows.

			\section*{Open Research Section}
			The software underlying this research is available on GitHub at \url{https://github.com/curian127/score-pff}. The repository includes: (1) the complete Python implementation of Score-PFF, including the particle flow filter framework, score network architectures (MLP, FNO, GNN), and training scripts; (2) experiment reproduction code for all results reported in this paper; (3) Jupyter notebooks for figure generation; and (4) pre-trained score network weights for both 40-dimensional and 1000-dimensional Lorenz-96 systems.
			All simulation data in this study are deterministically generated from the Lorenz-96 model (Eq.~\ref{eq:l96}) using the provided scripts. The 40-dimensional training dataset (1 million samples from 500 trajectories) and the 1000-dimensional training dataset (100,000 samples from 50 trajectories) can be reproduced by running generate\_training\_data.py with the random seeds specified in the repository documentation. Assimilation experiment outputs (ensemble trajectories, RMSE time series) can be reproduced by running the corresponding experiment scripts with documented hyperparameters.
			Key dependencies include Python 3.9, PyTorch 2.7.1, NumPy 1.24, and SciPy 1.10. The 1000-dimensional experiments were conducted on an Intel Xeon Platinum 8276 processor with 256GB system memory.

			\acknowledgments
			The authors acknowledge the use of Kimi K2.6 (Moonshot AI) for assistance with plotting code generation and language editing. The scientific content, methodology design, data analysis, and interpretations are solely the authors' own.
			This study was supported by the National Natural Science Foundation of China (42450178), and grants from the National Key Research and Development Program under contract No. 2025YFF0517203.

				%
				%
				
				\bibliography{mybib}

				%
				%
				%
				%
				%

			\end{document}